\documentclass[prb,twocolumn,showpacs,preprintnumbers,amsmath,amssymb]{revtex4}
\usepackage{graphicx}
\usepackage{dcolumn}
\usepackage{epsfig}

\begin{document}

\title{Upper critical field of electron-doped Pr$_{2-x}$Ce$_{x}$CuO$_{4-\delta}$ in parallel magnetic fields}

\author{Pengcheng Li$^1$}
\author{F. F. Balakirev$^2$}
\author{R. L. Greene$^1$}
\affiliation{$^1$Center for Superconductivity Research and
Department of Physics, University of Maryland, College Park,
Maryland 20742-4111 \\$^2$NHMFL, Los Alamos National Laboratory,
Los Alamos, NM 87545} \

\date{\today}

\begin{abstract}
We report a systematic study of the resistive superconducting
transition in the electron-doped cuprates
Pr$_{2-x}$Ce$_{x}$CuO$_{4-\delta}$ down to 1.5 K for magnetic
field up to 58 T applied parallel to the conducting ab-planes. We
find that the zero temperature parallel critical field
(H$_{c2\parallel ab}$(0)) exceeds 58 T for the underdoped and
optimally-doped films. For the overdoped films, 58 T is sufficient
to suppress the superconductivity. We also find that the Zeeman
energy $\mu_B$H$_{c2\parallel ab}$(0) reaches the superconducting
gap ($\triangle_0$), i.e. $\mu_B$H$_{c2\parallel ab}(0)\simeq
\triangle_0$, for all the dopings, strongly suggesting that the
parallel critical field is determined by the Pauli paramagnetic
limit in electron-doped cuprates.
\end{abstract}

\maketitle

The upper critical field H$_{c2}$ is a crucial parameter for
high-T$_c$ superconductors. It provides important information
about the superconducting (SC) parameters, such as coherence
length, SC gap, etc.\cite{Tinkhambook} In past years, numerous
transport experiments\cite{Ando, Fournier} on high-T$_c$ cuprates
in the H$\perp ab$ configuration have been reported and the
H$_{c2}$-T diagrams have been established. A positive curvature in
both cases was observed from the resistivity measurements, which
is in contradiction to the expected low temperature saturation in
the Werthamer-Helfand-Hohenberg (WHH) theory.\cite{WHH} The most
likely reason for this is that the complicated H-T phase diagram
of high-T$_c$ superconductors includes a broad region of a vortex
liquid state and strong SC fluctuations.\cite{Wangyy} These
properties are detrimental to the determination of H$_{c2}$ from
resistivity measurements. Recent high-field Nernst effect
measurements\cite{Wangyy} in hole-doped cuprates revealed a
different H-T diagram when H$_{c2}$ is determined by a loss of
vorticity. A significant increase of H$_{c2}$ and an extrapolation
of H$_{c2}$(T) to well above T$_c$ were found. This observation
was explained by the existence of a non-vanishing pairing
amplitude well above T$_c$, while long range phase coherence
emerges only at T$_c$. H$_{c2}$ could then be a measure of the
onset of pairing amplitude.

Most of the H$_{c2}$ results obtained so far on the cuprate
superconductors are in the H$\perp$ab configuration. The strong
anisotropy, which would result in a much higher H$_{c2}$ for
magnetic field parallel to the conducting plane (ab-plane), and
the limitation of laboratory accessible magnetic fields makes the
H$_{c2\parallel ab}$ determination impossible for most of the
cuprates. Nevertheless, a few H$_{c2\parallel ab}$ data have been
reported.\cite{Brien,Sekitani,Dzurak, Vedenveev} An early
work\cite{Welp} that predicted H$_{c2\parallel ab}(T=0)$ for
YBa$_2$Cu$_3$O$_{7-\delta}$ based the initial slope, $-dH_{c2}/dT$
near T$_c$, was shown to be an overestimation by recent
measurements.\cite{Brien,Sekitani} The reason for this is that WHH
theory only accounts for the orbital pair breaking, but in the
H$\parallel$ab orientation, the Pauli spin pair breaking effect
could also be important. In fact, a recent
measurement\cite{Vedeneev2} on an underdoped
Bi$_2$Sr$_2$CuO$_{6+\delta}$ in a pulsed magnetic field up to 52 T
found that the Pauli paramagnetic limit could explain the H$_{c2}$
for field parallel to the conducting layers.

Compared to the hole-doped cuprates, the electron-doped are
distinctive for having a much lower H$_{c2\perp
ab}$.\cite{Fournier} This implies a larger in-plane coherence
length, and thus a smaller orbital critical field for H parallel
to CuO$_2$ planes is expected. In addition, Nernst effect
measurements have shown that electron-doped cuprates have much
weaker SC fluctuations\cite{Hamza} compared to the hole-doped. In
this paper, we present systematic parallel critical field
measurements in the electron-doped
Pr$_{2-x}$Ce$_x$CuO$_{4-\delta}$ (PCCO) for doping (\textit{x})
throughout the SC region and establish the H$_{c2\parallel ab}$-T
phase diagram. We find that the low temperature parallel critical
field is large (above 58 T at 4 K) for the underdoped and
optimally doped films, while it is below 58 T for the overdoped
films.  We also find that the Zeeman splitting energy
$\mu_BH_{c2\parallel ab}$ approaches the SC gap. Therefore, we
conclude that the paramagnetic limit is the cause of the
suppression of superconductivity in the H$\parallel$ab
configuration.

Five PCCO films with various doping (\textit{x}=0.13, 0.15, 0.16,
0.17, 0.19) with thickness about 2500 \AA\ were fabricated by
pulsed laser deposition on SrTiO$_3$ substrates.\cite{Maiser}
Since the oxygen content has an influence on both the SC and
normal state properties of the material,\cite{Jiang} we optimized
the annealing process for each Ce concentration. The sharp
transition and low residual resistivity are similar to our
previous report,\cite{Yoram} which implies the high quality and
well-defined doping and oxygen homogeneity of our films.
Photolithography and ion-mill techniques were used to pattern the
films into a standard six-probe Hall bar. Parallel field
resistivity measurements were carried out using a 60 T pulsed
magnetic field at the National High Magnetic Field Lab (NHMFL) in
Los Alamos. Resistivity data traces were recorded on a computer
using a high-resolution low-noise synchronous lock-in technique
developed at NHMFL. The films were carefully aligned to ensure a
parallel field (within $\pm1^0$ with respect to the ab-plane) and
we found no signs of eddy current heating in the data.

\begin{figure}
\centerline{\epsfig{file=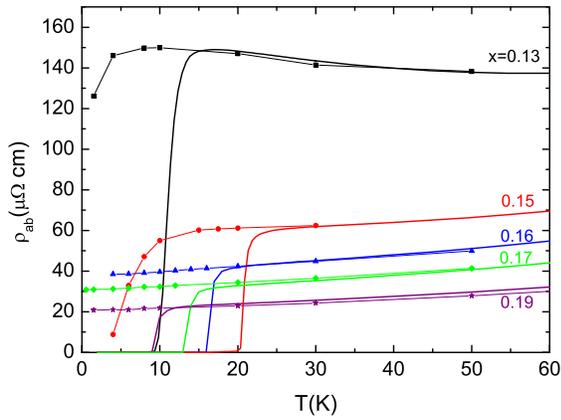,clip=,silent=,width=3in}}
\caption{(color online). In-plane resistivity versus temperature
in zero-field (solid lines) and H=58 T applied parallel to the
ab-planes (filled symbols) in PCCO films with various Ce
concentration.} \label{Fig1}
\end{figure}

\begin{figure}
\centerline{\epsfig{file=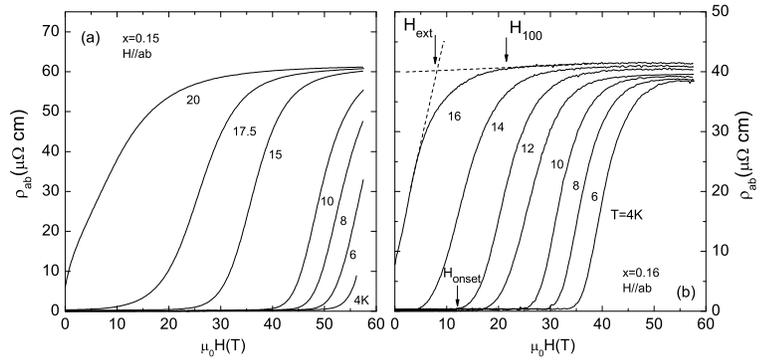,clip=,silent=,width=4in}}
\caption{In-plane resistivity versus magnetic field for
H$\parallel$ab-plane for (a) \textit{x}=0.15 (T$_c$=21.3 K) and
(b) \textit{x}=0.16 (T$_c$=16.9 K).}\label{Fig2}
\end{figure}

Fig.~\ref{Fig1} shows the in-plane resistivity ($\rho_{ab}$)
versus temperature in zero field and in 58 T for H$\parallel$ab
for all the films. The zero field transition temperatures are 10.8
K, 21.3 K, 16.9 K, 14 K, and 10.4 K for \textit{x}=0.13, 0.15,
0.16, 0.17 and 0.19 respectively. In the H$\perp$ab field
orientation, a field of order H$\leq$10 T is enough to suppress
the superconductivity, similar to previously work.\cite{Fournier}
However, when the field is aligned in the ab-plane, the
superconductivity is not completely destroyed in the underdoped
\textit{x}=0.13 and optimally doped \textit{x}=0.15 films even at
58 T, as seen in Fig.~\ref{Fig1}. In Fig.~\ref{Fig2} we show
$\rho_{ab}$(H) for H parallel to the ab-plane for the films
\textit{x}=0.15 and 0.16. Apparently, the normal state can not be
completely recovered in the optimally doped $x$=0.15 for T$\leq$10
K. However, for the overdoped film $x\geq$0.16, 58 T is sufficient
to destroy the superconductivity even at the lowest temperature
(1.5 K) measured. Compared to the H$\perp$ab
geometry,\cite{Fournier} a broader transition in $\rho_{ab}(H)$ is
observed for the parallel field orientation. A similar behavior
was found for the other dopings (not shown).

\begin{figure}
\centerline{\epsfig{file=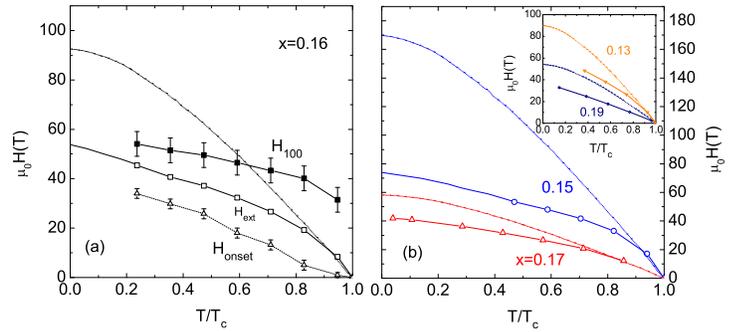,clip=,silent=,width=3.8in}}
\caption{(color online). (a) Resistive characteristic fields
H$_{onset}$, H$_{ext}$ and H$_{100}$ for H$\parallel$ab as a
function of reduced temperature T/T$_c$ for \textit{x}=0.16, (b)
H$_{ext}$ versus T/T$_c$ for \textit{x}=0.15 and 0.17. Inset shows
the data for \textit{x}=0.13 and 0.19. Dotted lines are fits to
the WHH theory.\cite{WHH} Solid lines are extrapolation based on a
smooth H(T) behavior.}\label{Fig3}
\end{figure}

From the $\rho_{ab}(H)$ traces in Fig.~\ref{Fig2}, we can
determine the resistive parallel critical field. However, the
choice of a criterion remains arbitrary, mainly because of the
curvature of the high-field flux-flow resistivity typical of all
high-T$_c$ superconductors. Following the schemes in the prior
work\cite{Ando, Fournier} as presented in Fig.~\ref{Fig2}(b), we
can determine the characteristic fields corresponding
approximately to the onset of flux flow (H$_{onset}$) and a higher
field corresponding to the complete recovery of the normal state
(H$_{100}$). In Fig.~\ref{Fig3}(a), we show H$_{onset}$ and
H$_{100}$ as a function of the reduced temperature (T/T$_c$) for
\textit{x}=0.16. The larger uncertainty of H$_{100}$ is marked
with larger error bars. In this figure, we also show the extracted
value (H$_{ext}$) at the extrapolation point of the flux-flow
region and the normal state asymptote. We find that H$_{ext}$ lies
between H$_{onset}$ and H$_{100}$ and it is close to the field
value determined from 90\% of the normal state resistivity. We
note that the H$_{ext}$ criterion has been regularly used as
representing an acceptable determination of H$_{c2}$ and we will
adopt H$_{ext}$ values as our estimate of H$_{c2\parallel ab}$.

\begin{figure}
\centerline{\epsfig{file=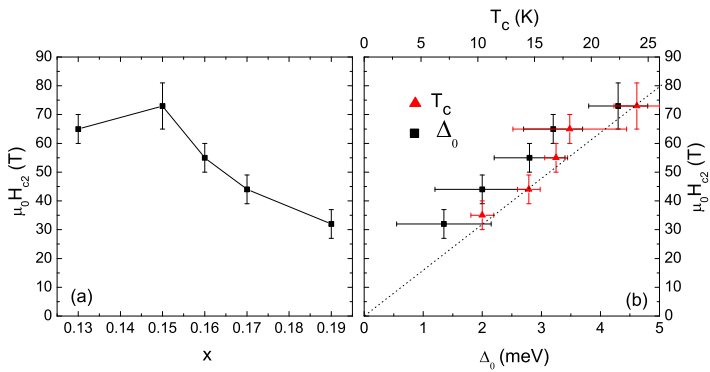,clip=,silent=,width=3.8in}}
\caption{(color online). (a) Doping dependence of extrapolated
H$_{c2\parallel ab}(0)$. (b) H$_{c2\parallel ab}(0)$ as a function
of T$_c$ and superconducting gap $\triangle_0$.}\label{Fig4}
\end{figure}

In Fig.~\ref{Fig3}(b), we plot the characteristic field H$_{ext}$
as a function of T/T$_c$ for the other films(we note that T$_c$ is
taken from resistivity in a procedure similar to H$_{ext}$). In
contrast to H$_{c2\perp ab}$(T),\cite{Fournier} no low temperature
divergence or positive curvature is observed in the H$\parallel$ab
configuration for most of the films. Although the low temperature
H$_{c2\parallel ab}$(T) behavior is unknown for \textit{x}=0.13
and 0.15 due to the limit of our field, from the overdoped films
data a saturation seems to emerge at low temperature, which is
similar to hole-doped cuprates.\cite{Sekitani, Vedeneev2} From the
H-T plots in Fig.~\ref{Fig3}, we can roughly extrapolate the
curves to get H$_{c2\parallel ab}(0)$ and its doping dependence is
shown in Fig.~\ref{Fig4}(a). A large zero temperature critical
field is found in the underdoped and optimally doped films, and a
dramatic decrease of H$_{c2\parallel ab}$(0) is observed for the
overdoped films. A similar trend was found in the doping
dependence of H$_{c2\perp ab}(0)$,\cite{Fournier, Mumtaz} both
H$_{c2\parallel ab}(0)$ and H$_{c2\perp ab}(0)$ decrease rapidly
in the overdoped region compared to the underdoped, although the
T$_c$ of underdoped films drops even faster.

We have established an experimental parallel field H-T diagram for
PCCO. Now let us compare our data with theory. For most
conventional superconductors, WHH theory can quantitatively
explain the temperature dependence of the upper critical field.
For the layered high-T$_c$ cuprates, in the H$\perp$ab
configuration, it is found that the upper critical field is in
good agreement with the WHH theory except for some unexplained low
temperature upward curvature.\cite{Vedeneev2} This implies that
the diamagnetic orbital effect dominates the paramagnetic spin
effect in the destruction of the superconductivity. In the
H$\parallel$ab geometry, we attempted to compare our data with WHH
theory (dotted lines in Fig.~\ref{Fig3}) by using the initial
slopes of the H-T plots. As shown in Fig.~\ref{Fig3}, for the
films near optimal doping (\textit{x}=0.15 and 0.16), we found
that WHH curves depart strongly from the experimental data at low
temperatures. To show this here, we take \textit{x}=0.15 as an
example. The zero temperature critical field obtained from the WHH
formula $H_{c2}(0)=0.693(-dH_{c2}/dT)\mid_{T=T_c}T_c$ is about 170
T(using the initial slope value at T$_c$,
$dH_{c2}/dT\mid_{T=T_c}$=-11.5 T/K), which is much larger than the
extrapolated value of 73 T. As seen in Fig.~\ref{Fig3}, the WHH
value of H$_{c2}$(0) is also larger than the experimental number
for \textit{x}=0.13 and 0.16. It appears that the WHH orbital
theory only sets the upper bound of H$_{c2}$(0) for these dopings.
However, we find that for the overdoped films, \textit{x}=0.17 and
0.19, the H$_{c2\parallel ab}$(0) values are close to the WHH
theoretical estimation.

For a layered superconductor, by neglecting the thickness of the
conducting layers, Klemm \textit{et al.}\cite{Klem} predicted that
the upper critical field would diverge for temperature below a
certain value T* where the out-of-plane coherence length $\xi_c$
decreases to the value $d/\sqrt{2}$(d is the distance between the
conducting layers) and a dimensional crossover from 3D to 2D would
occur at low temperature. The critical magnetic field to decouple
the layers at T* was predicted to be H$_c$=$\phi_0/d^2\gamma$
($\gamma=H_{c2\parallel ab}/H_{c2\perp ab}$). Experimentally, the
low temperature saturation in the H-T phase diagram for
H$\parallel$ab is contrary to this prediction and no trace of a
dimensional crossover is observed. The predicted H$_c$, which is
about 765 T for \textit{x}=0.15 (\textit{d}=6 \AA\ and
$\gamma\sim$8, a similar number is found for the other dopings),
is also very large. By considering the thickness (\textit{t}) of
the conducting layers, it has been found\cite{Tinkham2, Vedeneev3}
that the parallel critical field can be rewritten as
H$_{c'}=\sqrt{3} \phi_0/\pi t \xi_{ab}$. From our perpendicular
critical field data,\cite{Fournier} we can get the in-plane
coherence length $\xi_{ab}$ via the Ginzburg-Landau equation
$H_{c2\perp ab}=\phi_0/2\pi\xi^2_{ab}$. Setting the corresponding
values of \textit{x}=0.15 (t=3 \AA\, $\xi_{ab}$(0)=60 \AA), we
find H$_{c'}$=582 T, which is still much higher than our measured
value.

We now discuss paramagnetic (Pauli) limitation of the parallel
critical field. In this case, the electron spins couple with the
applied field and when the spin Zeeman energy reaches the
pairbreaking energy, the Cooper pair singlet state is destroyed.
An early theory by Clogston and Chandrasekhar\cite{Clogston}
estimated the paramagnetic limit based on the isotropic BCS theory
and predicted the Pauli paramagnetic limit $H_P=\triangle_0/\mu_B
\sqrt 2$. Under the assumption $2\bigtriangleup_0=3.5k_B T_c$, we
have $H_P(0)=1.84T_c\frac{T}{K}$. Applying this to our
\textit{x}=0.15 doping (T$_c$=21.3 K), we get H$_P(0)$=39 T. This
is much smaller than our experimental value of 73 T. If we take
$\bigtriangleup_0$=4.3 meV (maximum gap value) from the optics
results,\cite{Mumtaz, Homes} then $H_P'(0)$=53 T. For the other
dopings, we find that the Clogston theory also underestimates the
measured values. This suggests that a simple BCS s-wave model for
the paramagnetic limit is not valid for PCCO. This is not
surprising since PCCO is believed to be a quasi two dimensional
d-wave superconductor. Recent work by Yang\cite{Yang} estimated
the paramagnetic limit for a d-wave superconductor in a purely 2D
system by only considering the coupling of the spins of the
electrons and the applied field and found that
$H_P(0)=0.56\bigtriangleup_0/\mu_B$. This is even smaller than the
s-wave case due to the existence of nodes in the gap function.

The experimental critical field often exceeds the theoretical
predictions for the Pauli limit, even in some conventional s-wave
superconductors. To explain this, some other possibilities were
introduced, such as spin-orbit coupling to impurities. It was
found that the spin-orbit scattering enhances the Pauli critical
field over the spin-only value for s-wave symmetry.\cite{WHH,
Klem} However, it has been shown\cite{Grimaldi} that the
spin-orbit interaction significantly lowers the critical field for
d-wave symmetry. Therefore, the enhancement of the parallel
critical field in PCCO is most unlikely caused by the spin-orbit
coupling.

Despite the discrepancy between theory and data, we find that our
extrapolated H$_{c2\parallel ab}$(0) can be scaled with both T$_c$
and SC gap $\triangle_0$. As seen in Fig.~\ref{Fig4}(b),
H$_{c2\parallel ab}$ is linearly proportional to T$_c$ and can be
written in a Zeeman-like way, i.e., $k_B
T_c=\frac{1}{4}g\mu_BH_{c2\parallel ab}(0)$, where $g$=2 is the
electronic $g$ factor, $\mu_B$ the Bohr magneton. This suggests
that the thermal energy at $T_c$ and the electronic Zeeman energy
at $H_{c2\parallel ab}(0)$ give the single energy scale required
to destroy the phase coherence. We note that, for underdoped
\textit{x}=0.13 and optimally-doped \textit{x}=0.15, due to the SC
fluctuation, we determined T$_c$ from the temperatures at which
the vortex Nernst effect disappears, which is 18 K and 24 K for
0.13 and 0.15, respectively. This temperature is slightly higher
than the resistive transition temperature.\cite{Hamza} For the
overdoped films, both tunneling\cite{Yoram2} and Nernst effect
measurements show that the fluctuation is much weaker, therefore,
T$_c$ can be reliably taken from resistivity measurement.
Meanwhile, if we compare the Zeeman energy and the maximum SC gap
values obtained from optics,\cite{Mumtaz, Homes} we find that
$g\mu_BH_{c2\parallel ab}(0)\simeq2\triangle_0$, i.e.
$\mu_BH_{c2\parallel ab}(0)/\triangle_0\simeq 1$, as shown in
Fig.~\ref{Fig4}. This strongly suggests that the magnetic Zeeman
energy reaches the SC gap, and thus the superconductivity is
destroyed. It has been shown that due to possible quantum
fluctuations, the superconductivity can be destroyed within a
Zeeman energy interval,\cite{Aleiner} $\frac{1}{2}\triangle\leq
\mu_BH_{c2\parallel ab}\leq 2\triangle$. Therefore, our results
strongly suggest the Pauli paramagnetic limit is responsible for
the high field depairing process.

Finally, it is worth mentioning that the SC gap to parallel
critical field ratio in some hole-doped cuprates was also found to
be roughly one.\cite{Brien,Vedeneev2} It seems that in the layered
quasi-2D cuprate superconductors, the parallel critical field is
universally determined by the paramagnetic limit, suggesting that
diamagnetic orbital pair-breaking effect is negligible compared to
the spin effect due to a much shorter out-of-plane coherence
length.

In summary, we measured H$_{c2\parallel ab}$ in electron-doped
cuprates Pr$_{2-x}$Ce$_x$CuO$_{4-\delta}$ from the underdoped to
the overdoped region. We found that the critical field anisotropy,
$H_{c2\parallel ab}/H_{c2\perp ab}$ is about 8. We also found that
the Zeeman energy $\mu_B H_{c2\parallel ab}$(0) reaches the
superconducting gap $\triangle_0$, which strongly suggests that
the Pauli paramagnetic limit is responsible for quenching
superconductivity in electron-doped cuprates for H parallel to the
CuO$_2$ planes.

PL and RLG acknowledge the support of NSF under Grant DMR-0352735.
The work in NHMFL is supported by NSF and DOE.

\end{document}